# Logical-Combinatorial Approaches in Dynamic Recognition Problems

Levon H. Aslanyan[1], Viktor V. Krasnoproshin[2], Vladimir V. Ryazanov[3], Hasmik A. Sahakyan[1]

[1]Institute for Informatics and Automation Problems of NAS RA
[2]Belarusian State University, Department of Information Management Systems
[3]Dorodnitsyn Computing Centre, Federal Research Center Computer Science and Control, RAS
e-mail: lasl@sci.am, krasnoproshin@bsu.by, rvvccas@mail.ru, hsahakyan@sci.am

**Abstract**

A pattern recognition scenario, where instead of object classification into the classes by the learning set, the algorithm aims to allocate all objects to the same, the so-called "normal" class, is the research objective. Given the learning set $L$; the class $K_0$ *is called* "normal", and the reminder $l$ classes $K_1, K_2, \ldots, K_l$ from the environment $K$ are "deviated". The classification algorithm is for a recurrent use in a "classification, action" format. Actions $\mathcal{A}_i$ are defined for each "deviated" class $K_i$. Applied to an object $x \in K_i$, the action delivers update $\mathcal{A}_i(x)$ of the object. The goal is in constructing a classification algorithm $\mathfrak{A}$ that applied repeatedly (small number of times) to the objects of $L$, moves the objects (correspondingly, the elements of $K$) to the "normal" class. In this way, the static recognition action is transferred to a dynamic domain.

This paper is continuing the discussion on the "normal" class classification problem, its theoretical postulations, possible use cases, and advantages of using logical-combinatorial approaches in solving these dynamic recognition problems. Some light relation to the topics like reinforcement learning, and recurrent neural networks are also provided.

## 1. Introduction

Pattern recognition as a cybernetical research direction has been formed since the 50's of the previous century [1,2]. Two sides of schools of Soviet recognizers, led by Yu. Zhuravlev [3] and V. Vapnik [4], have consecutively become the leading force of this research domain worldwide and the boom of current machine learning and artificial intelligence developments. Now the theory is classical [5], with new dramatical developments concerned to the so called Deep Neural theory [6], which is mostly machine-oriented. Whilst traditional theories are oriented in constructing tractable and interpretable recognition theories [7-11], the deep theory is mostly computation-oriented. The core idea of recognition is in weak learning resource, but the deep learning operates with very large learning collections, so it is possible, theoretically, to derive the necessary interpretable knowledge but computational cost is becoming higher in these models.

The mainstream of classification technique includes models, such as PAC and logic-combinatorial, as well as algorithms such as boosting, bugging, SVM, kernel-based, and others. All deep learning techniques are based on neural networks. Along with the indicated popular and dominant approaches, both in classical theory and in deep theories, there are separate methods and algorithms that are aimed at solving individual non-standard problems, as well as problems with specific restrictions. This work is aimed at analyzing just such situations. Here are some examples.

The problem of classification of one class is considered in [12]. The set of training elements of this task presents examples of objects satisfying a certain class property, and there are no counter examples. The solution of the problem involves the step of construction of such a shell that borders all elements of the training set and does not contain outliers. The solution could be simply a convex hull or an isoperimetric hull [13]. In Boolean domain, the solution can be the reduced disjunctive normal form [14] and its extension [15]. A different scheme with one class classification is the algebraic, spectral algorithm for row and column weights in object characterization tables [16].

In contrast to one class case, there is a large number of publications devoted to the problem of classification with a very large number of classes, hundreds and thousands [17-19]. The solution is logic-combinatorial, through constructing a binary code table, the columns of which represent dichotomies of classes, with rows representing a clustering of all classes.

As a third example of non-standard pattern recognition, the High Dimensional Low Sample Size (HDLSS) data analysis paradigm may be mentioned. Learning table is very long, gigabytes for problems from genomics. There may even be a few classes, , but there are so many elementary classifiers, that without additional constructions or additional knowledge it is impossible to differentiate them. A logic-combinatorial approach to this problem is presented in [20].

Occasionally, the logic-combinatorial approach appears in all our non-standard examples. This method was initiated by early papers: [14] in terms of disjunctive normal forms, [21] in terms of tests and binary matrices, [22] in terms of Boolean expression, and [23] in terms of voting and similarity calculation. Later it turned out that all these schemes are cross interpretable [24].

This paper will introduce and analyze one more specific postulation/problem in pattern recognition [25]. The problem is a process of recurrent application of classification algorithms intending to move all objects to the one of the classes called "normal". For a given algorithm it is to find a way to estimate its work on classification to normal. And when an appropriate learning set is available, the problem is in constructing a well-optimized algorithm to solve the defined problem. We will start with the necessary definitions. After that we consider interpretations in terms of Markov Decision Processes and Reinforcement Learning. After this, a novel approach with logical-combinatorial algorithms is presented and discussed.

## 2. Problem Definition

*Classification-action recursion (CAR)*. The standard definition of a pattern recognition problem considers $n$ features, disjoint classes $K_1, K_2, \ldots, K_l$ of objects from some environment $K$ characterized by these features, and an $m$ object learning set $L = \{x_1, x_2, \ldots, x_m\}$, where $L \cap K_i, i = 1,2, \ldots, l$ is the share of the $i$-th class in the learning set. The goal is to create a classification algorithm $\mathfrak{A}$ based on the learning set, which classifies objects in the environment $K$ as accurate as possible. Additional information about the classes and classification is a benefit of the model.

We consider a principally different version of this pattern recognition problem. Here it is assumed that one of the given classes is "normal", let it be denoted by $K_0$, and all the other classes are the so called "deviated" classes. Also, we are given a finite set $\mathcal{A}$ of actions/functions $a$, that being applied to the objects $x \in K$, deliver their functional updates $a(x)$, keeping them in the same environment $K$. In the simplest case, we assume that a certain action $a_i \in \mathcal{A}$ is attached to every "deviated" class $K_i$ (the $i$-th class action); and being applied to an arbitrary object $x \in K_i$, delivers its update $a_i(x)$. $a_i(x)$ may be allocated to anyone of the classes and it is not necessary that this is the same initial class $K_i$, or some unique class for all objects of $K_i$. The goal is in constructing a classification algorithm $\mathfrak{A}$, that applied repeatedly (small number of times) to the objects of $L$ (correspondingly, to the elements of $K$) moves these objects to the "normal" class.

Thus, the process is as follows: Algorithm $\mathfrak{A}$ is applied repeatedly to the elements of learning set $L$ (in the first step algorithm $\mathfrak{A}$ is applied to the elements of $L$, and in the next steps - to their updates by the set of class actions). If after a current $k$-th repetition/application of the algorithm there still remains an object $x \in L$, or an object appeared during the process, which is classified to a class other than $K_0$ (say, to some deviated class $K_i$), then at the next $(k + 1)$-th repetition of $\mathfrak{A}$, the action $a_i$, attached to the class $K_i$ is applied on $x$, updating the learning set labels in this way.

*Use-case of treatment regime (TR).* Consider a generative application scenario of the described framework from the medical domain – consider the Dynamic correction approach of the patient's treatment regime [26]. Here, "classification operation" means that the current diagnose is obtained by the medical doctor for any object of classes $1 \div k$. There is no reason to apply classification to the elements of "normal" class because its elements represent healthy cases. Recall that each of the classes $K_i$ is $1 - 1$ related to their actions $a_i$, and in this case, $a_i$ is the treatment action for class $K_i$. It is evident that the overall goal is to bring the patients, after several treatment stages, to the "normal" class. Two different subcases of this use case will be considered. At first we suppose that the records and observations of only one particular doctor are available. In this case, we aim at estimating the effectiveness of the diagnostic approaches applied by the doctor. In the second scenario, we suppose that we are given larger information of a set of doctors and we try to determine the optimal way of diagnoses to achieve the best allocation result to the "normal" class. From an algorithmic point of view, this is a kind of inversere cognition problem. Ordinary recognition aims at mimics of the one-step classification actions. Here, for an algorithm that we apply recurrently, we need to guess all ancestors that will be mapped onto the predefined class. Moreover, it is necessary to generate an algorithm with the set of ancestors larger than the learning set.

As we mentioned, two main classification scenarios and corresponding problems will be considered:

**Scenario 1**. The basic information/knowledge available of this scenario is defined in the special form of the learning set $L$ of the classification problem. Although the class actions are automatically applied to the elements of the deviated classes, and each reapplication of the algorithm may work with the updated objects, however, we are given neither the details of this information, nor the updates themselves. We only suppose that it is empirically accepted/supposed that the set $L$ is obtained/recorded in practice by a witness in the form of an object-class-label triple, and the objects of $K$ tend to be classified to the class $K_0$ in a few repeated applications of the algorithm $\mathfrak{A}$ is a supposition, but this needs to be verified.

In its complete form, the learning set $L$ is a data flow. The considered objects $x$ have their identifiers $I_x$, which is a many-to-one mapping in a time interval, many classes, one identifier. Object $x$, after operated by the algorithm $\mathfrak{A}$, changes its time stamp. Initial time stamp is the time $t_0$ of the first appearance of the object in the algorithm $\mathfrak{A}$. After classification and action is applied to $x$, $x$ accepts the modified value $x^{(1)}$ and the new time stamp $t_1$ with $t_0 < t_1$. In this way, objects travel through the classes forming the so called traces, $t_0, t_1, \ldots, t_k$ and $x = x^{(0)}, x^{(1)}, \ldots, x^{(k)}$. The basic objective is to insure that the end points of traces belong to the class "normal". In this Scenario 1, we have a bystander, witness, who cannot see the timestamp and identifiers. In this limited information, the problem formed will try to verify whether the strategy of algorithm $\mathfrak{A}$ is supportive to classification to the class "normal". Consider two particular issues here:

**Problem 1.1**. Assess the compliance and validate the empirical classification algorithm $\mathfrak{A}$ "to the class normal" based on the deterministic learning set $L$.

In terms of *TR*, an individual doctor, after a particular diagnosis with classification to the deviated class $K_i$, applies its unique action for that case. So, the class-to-class transition process is deterministic. The model relation over the classes is a partially ordering relation – if action $a$ of a class $x$ is applied, then the deterministic transition is to class $y$, and we code this by $y \leq x$. The partially ordered set (poset) of classes may have several minimal elements. Our goal is to check, if the model has a minimum, so that it is the "normal class". The brute force combinatorial approach may do this work. If at systematic or logical level, then it is to check several facts: L is a poset, L is connected, L has a unique minimal element. Poset relations are the reflexivity $aRa$, anti-symmetry $aRb \,\&\, bRa \Leftrightarrow a = b$, and transitivity $aRb \,\&\, bRc \Rightarrow aRc$. Complexity of these checks is quadratic by number of classes. The check for a minimum may use the following: if in a poset $L$ there is a unique minimal element, and, for every subset $L' \subseteq L$ there is an element $m$ such that there is no $s' \in L$ with $s' < m$ and there exists an element $s' \in L$ such that $m \leq s'$, then $S$ has a minimum element. And the combinatorial check might be by construction of the Hasse diagram of $L$ by the following steps. We put the class "normal" at the bottom of the page/diagram. At a level above the "normal", we put all those classes (their labels) that have direct transition/link to the lower level vertices (at the first step this vertex of class "normal"). These links can follow immediately the lower level, but not the further levels. This procedure is continued until its expiration, and we denote the final tree by $\Xi$. The only condition that there are no classes outside the diagram, is satisfactory to say that the check of Problem 1.1 is tested positively.

**Problem 1.2**. Assess the compliance and validate the empirical classification algorithm $\mathfrak{A}$ "to the class normal" based on the stochastic learning set $L$.

The *TR* interpretation of Problem 1.2 is also possible. This is when information is collected from a set of doctors. Each doctor $j$, faced with a case of a deviated class $K_i$, applies his action $a_{ji}$ transitioning to some class $K_{ji}$. Two types of branching are possible here. The first one is when it is possible to use different actions at the same state, and the second one is when, after exposure, the object can fall into different classes. We will restrict ourselves to considering the second case. The existing theoretical model of such a behaviour is the model of Markov decision processes (*MDP*). *MDP* is a model with 4-tuple $(S, A, P_a, R_a)$, where $S$ is the set of *States* (diagnoses in *TR*), $A$ is the set of *Actions* (prescriptions, a set of actions may be different around the different states), $P_a(s, s')$, the probability that a defined action $a$ in state $s$ applied at discrete time $t$ leading to the state $s'$ at next time stamp $t + 1$ (or one action is applied but the next states may be different), and $R_a(s, s')$ is the so called *Reward* from this transition (achieved or approached to the normal class). $R_a(s, s')$ can be associated to the distance of a state vertex on the tree $\Xi$ to the root vertex that corresponds to "normal". $R_a(s, s')$ forms the so called value function of MDP, which is an object to be optimized. We do not need to enter this theory, which is well known but some notes are necessary. The *MDP* objective is to achieve an optimal decision/policy $\pi(s)$ at the state $s$. $\pi(s)$ is based on the analysis of the learning set. Main optimization ideas include the Bellman equations and value function analysis. The fundamental classes of methods for solving finite Markov decision problems are dynamic programming, Monte Carlo methods, and temporal-difference learning. Each of them delivers a policy $\pi(s)$ based on $(S, A, P_a, R_a)$. On the contrary, within the framework of Problem 1.2, $\pi(s)$ is observable. Being interested in testing/evaluating this policy $\pi(s)$, it is to generate the optimal policy by *MDP* and compare it with the observed one.

**Scenario 2**. The learning set $L$ is updated after each reapplication of the classification algorithm $\mathfrak{A}$, according to the class actions results/updates. This scenario, as we see, collects the complete information from environment, recording current characteristics of objects, their classification with one of several actions linked to this class, time stamps, identifiers. Using time stamps and identifiers, it is easy to construct the chain of passage by classes. At the horizontal, time free level, there is still additional information over Scenario 1 in the form of identifiers that allows us to separate the information about the individual object. In fact, Scenario 2 may have many subcases to be carefully defined and studied.

**Problem 2.1**. Deterministic optimized classification algorithm $\mathfrak{A}$ including additional information on the objects and updates.

In Scenario 2, the learning set $L$ is updated/extended after each reapplication of the classification algorithm. In this case, the object ID is recorded in all steps, and this provides a follow up mechanism through the recurrent classification processes. Let $x \in L$, $x \in K_i$, and let some empirical treatment of $x$ be applied. That is, $x$ is classified to the class $K_i$; after that, the action $a_i$ (the $i$-th class action) is applied, and as a result $x$ is modified into the $y$: $a_i(x) = y \in K_j$. In this way, chains are appearing in the course of repeated classifications, and some of these chains lead to the class $K_0$.

In a formal description, the learning set $L$ is represented by a linkage graph $G$ with the vertex set $V$ corresponding to the learning set elements, the set $V$ and the graph itself extended during the process, and its directed edges compose the set $E$, labeled by actions, and connecting the pairs of learning set elements. An edge may or may not have a weight. In this manner, the graph $G$ provides valuable information for checking the model validity, obtaining realistic information about the applied problems. The graph-theoretical problems that appear here helping to check the system, are well known and investigated theoretically. And nowadays research tendencies provide innovative applied approaches and analytics to the graph connectivity, expansion characterization, effective distances and other topics, in terms of sparse symmetric diagonally dominant matrix computational theory [27] that gives an acceptable implementation to the Linkage graph model algorithms and the corresponding software platforms.

**Problem 2.2**. Optimized classification algorithm $\mathfrak{A}$ including additional information on the objects and updates.

This form of the problem *TR*, in its general form, refers to one of the novel conceptual directions of the machine learning known as Reinforcement Learning (*RL*). The goal of *RL* is an interaction model of an agent with the environment by obtaining and analyzing the so called rewards of environment to these actions. The goal is to create an optimal acting agent (a doctor in our use case *TR*) for successful interacting with the environment (with the set of patients).

Action in our representation is learned to classify all objects to the unique "normal" class. So, when the class label is different from "normal", the action gets a penalty that depends on the distance between the detected class and the "normal" class. This approach can also be presented in a form of recurrent neural network model [31]. A weaker relation is with the known *inverse classification model* [28, 29], which is an analysis of the space of features, and the groups of features, providing a better one-class classification. No other systematically studied and related areas are known. The mentioned technique is tightly related to the backpropagation approach. Backpropagation has a very broad scope, and the "normal" class classification discipline appears as an inverse recognition problem. One step back gives the area that will/may be mapped to the class "normal". It is to differ the objects that necessarily will be classified to "normal" ($\forall$), the objects that have never been mapped to "normal" ($\emptyset$), and others that are classified to classes according to some probabilistic distributions, and the class "normal" is among these classes ($\exists$). Next step back accepts a similar picture of classification. Our goal is to determine all objects always classified to "normal", and those will be allocated to "normal" at least one time. And, of course, we are interested to know the frequencies of these allocations. In RL the way is through the *MDP*, dynamic programming, Bellman equations, and policy optimization. Our technique to achieve this information is the LCPR model and algorithms.

## 3. Proposed Methods and Solutions

In this section, in a compact form, we start with basic definitions from the logic-combinatorial pattern recognition theory. This theory will be used in solving Problems 1-2 and other similar problems. Consider a typical case recognition problem with $n$ features, $l$ disjoint classes $K_1, K_2, \ldots, K_l$ from an environment $K$ and an $m$ object learning set $L = \{x_1, x_2, \ldots, x_m\}$. $L_i = L \cap K_i, i = 1,2,\ldots,l$ denotes the share of the $i$-th class of the general learning set, that we suppose, is not empty. Objects are coded and are identical to their descriptions in the form of a vector of feature values: $x = (x_1, x_2, \ldots, x_n)$. For simplicity, we assume that $x_i \in R$, $i = 1,2,\cdots,n$.

Let us define the following set of elementary predicates, parametrically dependent on support sets $\omega_1, \omega_2 \subseteq \{1,2,\ldots,n\}, |\omega_1| = k_1, |\omega_2| = k_2$ and vectors $c_1 \in R^{k_1}$ and $c_2 \in R^{k_2}$. Below we use the notation $(x \leq a) = \begin{cases} 1, x \leq a, \\ 0, otherwise \end{cases}$.

**<u>Definition 1.</u>** ([15])
The predicate $P^{\omega_1,k_1,\omega_2,k_2}(x) = \bigwedge_{j \in \omega_1}(c_{1,j} \leq x_j) \bigwedge_{j \in \omega_2}(x_j \leq c_{2,j})$ is called a logical dependency (LD, geometrically based on a parallelotope) of the class $K_i$, if
1. $\exists x_t \in L_i: P^{\omega_1,k_1,\omega_2,k_2}(x_t) = 1$,
2. $\forall x_t \notin L_i: P^{\omega_1,k_1,\omega_2,k_2}(x_t) = 0$,
3. $P^{\omega_1,k_1,\omega_2,k_2}(x) = extr(F(P^{\omega_1,k_1,\omega_2,k_2}(x)))$,

where $F$ is the predicate quality criterion.

It is clear that the defined predicate geometrically presents a parallelotope; and then the function $F$ requires finding the local maximization of LDs in the domain. We will denote the set of all LDs of the $i$-th class of the given problem by $P_{L_i}$, and the set of all LDs of all classes by $P_L$. The predicate, satisfying only the first two constraints, is called admissible. We also accept the approximate predicates with a limited level of violation of the condition 2.

LD is the base element of the logic-combinatorial pattern recognition (LCPR) theory. The initial idea with LD appeared in [21]. Multi-parametric voting algorithms over the LD were introduced in [23,14] and obtained a complete analytics for LD with binary features. Here the predicates are maximal intervals/subcubes of the partially defined Boolean function, and the set of predicates is given by the reduced disjunctive normal forms of these functions.

*Principle of indeterminedness areas.* Let $L_1$ and $L_2$ be disjoint collections of classes, and let $P_1$ and $P_2$ be the LDs of these classes, correspondingly. $P_1$ and $P_2$ cover areas that may intersect, giving the indeterminedness area of LCPR.
- Areas dominated by $P_1$ and $P_2$ might be allocated to $L_1$ and $L_2$, correspondingly.
- Besides the two dominating and one intersection, there is no more areas because this will violate the condition 2. of Definition 1.
- Enlargement of $L_i$ is decreasing the indeterminedness domain.

*Search for LDs.* The transparent way is through the pricking the space by the finitely many class elements. This is exactly the scheme of constructing the reduced disjunctive normal form of Boolean function by consecutive application of zero assignments. Another way is the reduction to an IL program.

**Definition 2.** ([3])
LCPR similarity measure of an object of recognition $x$ to the class $K_i$ is:
$$\Gamma_i(x) = \frac{1}{|P_L|} \sum_{P^{\omega_1,k_1,\omega_2,k_2} \in P_L} P^{\omega_1,k_1,\omega_2,k_2}(x).$$

In short description, the LCPR stands out by:
- effective measure of similarity,
- proven separation of classes,
- multi-parametric optimization over large sets of recognition algorithms,
- correction of sets of algorithms providing correct recognition for all objects recognized by at least one individual recognizer, and other properties.

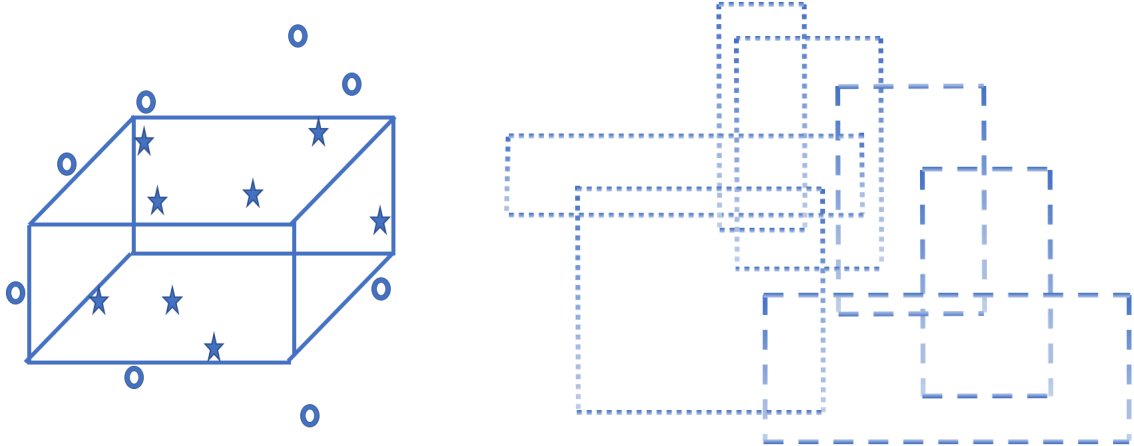

Left picture presents maximal LD, defined by 2 classes, "normal" and deviated in our case. Right side picture presents LDs for "normal" area with the LDs of counter-class areas.

Advantages of using the LCPR in solving dynamic recognition problems

In a general recognition algorithm $\mathfrak{A}$ by the learning set $L$ there is no visible idea on how to follow with repeated classifications, with the chains of classifications. However, the situation is different with the LCPR, because here it is possible to apply a backward reconstruction procedure of logical dependencies. At first, the set of LDs for the class $K_0$ is constructed by $L$. To do this, it is to form the counter-class that in a simple case can be the union of all deviated classes. A more adequate decision might be the selection by the rooted tree $\Xi$ that we constructed in Section 3. If the tree height is $h$, then, for example, taking $h/2$ we may compose the counter-class by all classes in a distance threshold $h/2$ and higher from the "normal". As it was mentioned, this is a set of parallelotopes in $R^n$. We suppose that all elements covered by these LDs create a new artificial class $K_*$, and one may now construct LDs defined by this class and by $L$. The Cartesian multiplication of the previous stage LDs is the way of creating new LDs. Continuing the process of LDs growing, in parallel we compare the covered volume of the object space with the size of $L$. Why is this possible technically?

Having only information of Scenario 1, we construct the partially ordered set $\Xi$ of classes as follows. The lower level, as we have mentioned, consists of one class $K_0$. At level 1, we place all those classes that have a direct link to the class $K_0$. Denote these classes by $S_1(K_0)$ and $S_1(S_1(K_0))$. The $S_1(S_1(K_0))$ together with $S_1(K_0)$ compose the second neighborhood $S_2(K_0)$ of $K_0$. An additional element of this model can be the threshold of links between the classes. To be a member of the neighborhood of $S_l(K_0)$, it is necessary that share of the links between the class and the set $S_l(K_0)$ is not less than the threshold. By these descriptions we see that LCPR might be applied to the set of classes of levels from 0 to some $m_1$ versus the classes of some level $m_2 > m_1$ and higher.

Implementation of this technique is not straightforward, it needs the knowledge gained on LCPR, as well as development of new approximate parallelotope-set type coverage approaches keeping the appearing complexities tractable.

It is worth mentioning that LCPR with LD provide the partial geometrical data structure, that helps not only with complexity controls, but also provides interpretability of results; and this is the known comparative benefit of all LCPR approaches.

The LCPR domain has been introduced and investigated for decades, resulting in hundreds of publications and scientific theses. Most investigated is the binary case. Here the reduced disjunctive normal form (RDNF) of partially defined Boolean functions is the analytical basis that helps to describe these classes of objects. In the simpler case of two classes, two RDNF of ordinary Boolean functions are considered. The first one is for the positive Boolean function that is true on the elements of "normal" class learning elements and the second one is for negation, for counter-class of this function. [3] shows that intersection of these two RDNF by LCPR will correspond to ($\exists$), while the positive intervals/subcubes will denote the ($\forall$) parts of the learning set. The ($\forall$) of positive ("normal" learning set) Boolean function represents the backpropagation of the "normal" class.

Next step to back from "normal", by the tree, is similar to the first step. Here new intervals/subcubes will be formed, and the core essence of procedure is the fact that all elements of the first step intervals/subcubes will be enlarged similarly, which draws to the Cartesian degrees of the intervals/subcubes. This is probably not easy computationally, but is a visible and interpretable analytics to inverse recognition procedures.

Multiclass extension is not difficult. It is to consider one-to-many classifications for all classes. This brings to a scheme consisting of $l + 1$ RDNFs. The rest of the process is similar to the two-class example. Of course, this is an initial interpretation of the inverse recognition model by the use of LCPR. The studies will be continued and implemented in practice. We evaluate this initial step taken over the new pattern recognition problem as important, because the applied model is ultimately practical, although being not yet formed and studied in its complete form.

## 4. Conclusion

More than 70 years of development of pattern recognition, which is now referred to by the term machine learning, has made it possible to formulate a solid set of models and technologies of both statistical and logical combinatorial nature. The variety is huge, both in the form of models and scenarios, as well as technologies and algorithms. The emergence of new tasks that have not yet been formed and not studied is not exclusion. One of such tasks - the procedure of assigning all objects to one fixed class by several consecutive steps of recognition is our objective. An applied use case problem of this type may be optimization of the course of treatment in medicine. This paper considers algorithms of pattern recognition of logical regularities in the context of solving the problem of assignment of objects to one fixed class. Only the initial analysis and research on this problem is characterized, its connection to the concepts of reinforcement learning and recurrent neural networks is indicated. Subsequent investigations of the problem will prove usefulness of the model in a number of applied problems. And especially, the related graph-theoretical and sparse linear algebra algorithms will be incorporated into the solutions.

## References

1. Бонгард М. М., Проблема узнавания, Москва, Физматгиз, 1967.
2. Rosenblatt F., The Perceptron: A Probabilistic Model for Information Storage and Organization in the Brain, Cornell Aeronautical Laboratory, Psychological Review, v65, No. 6, pp. 386–408, 1958.
3. Журавлев. Ю. И., Избранные научные работы, Магистр, Москва, 1998.
4. Вапник В. Н., Червоненкис А. Я., Теория распознавания образов, Москва, Наука, 1974.
5. Mohri M., Rostamizadeh A., Talwalkar A., Foundations of Machine Learning, The MIT Press, 414p., 2012.
6. Goodfellow I., Bengio Y., Courville A., Deep Learning (Adaptive Computation and Machine Learning Series), MIT Press, 2016.
7. Аркадьев А. Г., Браверман Э. М., Обучение машины распознаванию образов, М., Наука, 1964.
8. Minsky M. L. and Papert S. A., Perceptrons: an introduction to computational geometry, Cambridge, MIT Press, 1969.
9. Novikoff A. B., On convergence proofs on perceptrons, Symposium on the Mathematical Theory of Automata, 12, 615–622, Polytechnic Institute of Brooklyn 1969.
10. L. Aslanyan, J. Castellanos, Logic based pattern recognition ontology content (1), Proceedings iTECH-06, Varna, Bulgaria, pp. 61-66, 2006.
11. L. Aslanyan, V. Ryazanov, Logic based pattern recognition ontology content (2), Information Theories and Applications, vol. 15, no. 4, pp. 314-318, 2008.
12. Leng Q., Qi H., Miao J., Zhu W., Su G., One-Class Classification with Extreme Learning Machine, Hindawi Publishing Corporation, Mathematical Problems in Engineering, Volume 2015, 11 pages.
13. Асланян Л. А., Дискретная изопериметрическая задача и смежные экстремальные задачи для дискретных пространств, Проблемы кибернетики, том 36, стр. 85-128, 1979.
14. Асланян Л. А., Об одном методе распознавания, основанном на разделении классов дизъюнктивными нормальными формами, Кибернетика, т. 5, стр. 103 -110, 1975.
15. Рязанов В. В., Логические закономерности в распознавании образов (параметрический подход), Журнал Вычислительной Математики и Математической Физики, т. 47, номер 10, стр. 1793-1808, 2007.
16. Васильев Ю. Л., Дмитриев А. Н., Спектральный подход к сравнению объектов,

охарактеризованных набором признаков, Доклады АН СССР, 1972, т. 206, номер 6, стр. 1309–1312.
17. Dietterich T., Bakiri G., Solving multiclass learning problems via error-correcting output codes. Journal of Artificial Intelligence Research, 2: 263–282, 1995.
18. Zhuravlev Yu. I., Ryazanov V. V., Aslanyan L. H., and Sahakyan H. A., On a Classification Method for a Large Number of Classes, Pattern Recognition and Image Analysis, 2019, ISSN 1054-6618, Vol. 29, No. 3, pp. 366–376.
19. Zhuravlev Yu. I., Ryazanov V. V., Ryazanov V. V., Aslanyan L. H., and Sahakyan H. A., Comparison of Different Dichotomous Classification Algorithms, Pattern Recognition and Image Analysis, 2020, Vol. 30, No. 3, pp. 303–314.
20. Arakelyan A, Aslanyan L, Boyajyan A. High-throughput Gene Expression Analysis Concepts and Applications. Sequence and Genome Analysis II – Bacteria, Viruses and Metabolic Pathways. ISBN: 978-1-480254-14-5. iConcept Press Ltd, USA , 2013, 71-95.
21. Дмитриев А. И., Журавлев Ю. И., Кренделев Ф. П., О математических принципах классификации предметов или явлений, Дискретный анализ, Новосибирск, ИМ СО АН СССР, 1966. вып. 7., С. 3-17.
22. Вайнцвайг М. Н., Алгоритм обучения распознаванию образов "кора", Алгоритмы обучения распознаванию образов, под ред. Вапник В. Н., Москва, Советское радио, 1973. стр. 110–116.
23. Журавлев Ю. И., Никифоров В. В., Алгоритмы распознавания, основанные на вычислении оценок, Кибернетика, 1971, н. 3.
24. Aslanyan L., Ryazanov V., Sahakyan H., Testor and Logic Separation in Pattern recognition, Mathematical Problems of Computer Science, vol. 44, pp. 33-41, 2015.
25. Aslanyan L., Ryazanov V., Sahakyan H., On logical-combinatorial supervised reinforcement learning, International Journal "Information Theories and Applications", Vol. 27, Number 1, pp. 40-51, 2020.
26. Zhang Z., Reinforcement learning in clinical medicine: a method to optimize dynamic treatment regime over time, Annals of Translational Medicine, 2019, 7(14):345.
27. Gary L. Miller, The Revolution in Graph Theoretic Optimization Problems, In Proceedings of the 27th ACM on Symposium on Parallelism in Algorithms and Architectures, SPAA 2015, Portland, OR, USA, June 13-15, 2015, pp. 181, 2015.
28. Neu G., Szepesvári C., Apprenticeship learning using inverse reinforcement learning and gradient methods, In: Proc. 23rd Conf. Uncertainty in Artificial Intelligence, (2007) pp. 295-302.
29. C. C. Aggarwal, C. Chen, and J. Han, The inverse classification problem, Journal of Computer Science and Technology, vol. 25, no. May, pp. 458-468, 2010.
30. Sutton R, Barto A., Re-Inforcement Learning: An Introduction, MIT Press, Cambridge, MA, 1988.
31. Gimenes V., Aslanyan L., Castellanos J., Ryazanov V., Distribution Function as Attractors for Recurrent Neural Networks, Pattern recognition and image analysis, vol. 11.3, 2001, pp. 492-497.

## Տրամաբանական-կոմբինատոր մեթոդներ դինամիկ ճանաչողության խնդիրներում

Լ.Հ. Ասլանյան[1], Վ.Վ. Կրասնոպռոշին[2], Վ.Վ.Ռյազանով[3], Հ.Ա. Սահակյան[1]

L. Aslanyan, V. Krasnoproshin, V. Ryazanov, H. Sahakyan 11

[1]ՀՀ ԳԱԱ Ինֆորմատիկայի և ավտոմատացման պրոբլեմների ինստիտուտ
[2]Belarusian State University, Department of Information Management Systems
[3]Dorodnitsyn Computing Centre, Federal Research Center Computer Science and Control, RAS
e-mail: lasl@sci.am, krasnoproshin@bsu.by, rvvccas@mail.ru, hsahakyan@sci.am

## Ամփոփում

Դիտարկվում է նոր և հետաքրքիր ընթացակարգ կերպարների վերծանման տիրույթում, որտեղ առարկաների ճշգրիտ դասակարգման (ըստ ուսուցման բազմության) փոխարեն դրվում է այլընտրանքային նպատակ՝ դասակարգել բոլոր առարկաները մինույն, այսպես կոչված, «նորմալ» դասին: Տրված է ուսուցման $L$ հավաքածու, և դասերի մեջ կա մեկ նորմալ դաս՝ $K_0$, և $l$ "շեղված" դասեր՝ $K_1, K_2, \cdots, K_l$, - որևէ $K$ միջավայրից: Ուսուցման պրոցեսը դինամիկ է՝ «դասակարգում, գործողություն» ձևաչափում հետևյալ ձևով. $K_i$, "շեղված" դասերից յուրաքանչյուրին կցված է $A_i$ գործողություն/ֆունկցիա, որը կիրառելով կամայական $x \in K_i$ առարկայի վրա, արդյունքում ստացվում է այդ առարկայի թարմացումը՝ $A_i(x)$, որը կրկին $K$ միջավայրից է: Արդյունքում, $A_i(x)$-ը կարող է դասակարգվել կամ "շեղված" դասերից որևէ մեկին (ներառյալ նույն $K_i$ դասը), կամ՝ «նորմալ» $K_0$ դասին: Նպատակը հետևյալն է. կառուցել դասակարգման $\mathfrak{A}$ ալգորիթմ, որը կրկնողաբար (փոքր թվով անգամներ) կիրառվելով $L$-ի առարկաների վրա, ի վերջո առարկաները (համապատասխանաբար, $K$ – ի տարբերը) տանում է «նորմալ» դասի: Այսպիսով, ստատիկ ճանաչողության գործողությունը տեղափոխվում է դինամիկ տիրույթ:

Աշխատանքը, ընդհանուր առմամբ, իրենից ներկայացնում է քննարկումներ նշված խնդրի շուրջ՝ տեսական պոստուլատներ, հնարավոր կիրառություններ և դինամիկ ճանաչման խնդիրների լուծման գործում տրամաբանական-կոմբինատոր մեթոդների օգտագործման առավելությունների բացահայտում:

Հաշվի է առնվում նաև առնչությունը այնպիսի թեմաների հետ, ինչպիսիք են՝ ամրապնդմամբ ուսուցումը, տրամաբանական-կոմբինատոր ճանաչողությունը, ռեկուրենտ նեյրոնային ցանցերը:

**Բանալի բառեր**՝ դասակարգում, տրամաբանական-կոմբինատոր մոտեցում, վերահսկվող ուսուցում՝ ամրապնդմամբ:

# Логико-комбинаторные методы в задачах динамического распознавания

Л.А. Асланян[1], В.В. Краснопрошин[2], В.В.Рязанов[3], А.А. Саакян[1]

[1]Институт проблем информатики и автоматизации НАН РА
[2]Belarusian State University, Department of Information Management Systems

[3]Dorodnitsyn Computing Centre, Federal Research Center Computer Science and Control, RAS
e-mail: lasl@sci.am, krasnoproshin@bsu.by, rvvccas@mail.ru, hsahakyan@sci.am

**Аннотация**

Рассматривается новая важная процедура в области распознавания образов, где вместо точной классификации объектов по обучающему набору, ставится альтернативная цель отнесения всех объектов к одному и тому же, к так называемому, «нормальному» классу. Задан обучающий набор $L$; среди классов есть один «нормальный» класс $K_0$, и $l$ «отклоненных» классов $K_1, K_2, \cdots, K_l$ в некоторой среде $K$. Процесс обучения является динамическим в формате «классификация, действие» следующим образом: определенное действие/функция $A_i$ прикрепляется к каждому из «отклоненных» классов $K_i$, так что применяя действие $A_i$ к произвольному объекту $x \in K_i$, его обновление $A_i(x)$ остается в среде $K$. В результате, $A_i(x)$ может быть отнесен либо к одному из отклоненных классов (включая тот же класс $K_i$), либо к «нормальному» классу $K_0$. Задача заключается в построении такого алгоритма классификации $\mathfrak{A}$, который, многократно (небольшое количество раз) применявшийся к объектам $L$, переводит объекты (соответственно, элементы $K$) в «нормальный» класс. Таким образом, статическое распознавание переносится в область динамического распознавания. Статья представляет обсуждение проблемы, ее теоретических постулатов, возможные применения на практике, а также выявление преимуществ использования логико-комбинаторных подходов при решении этих задач динамического распознавания. Учитывается отношение к таким темам, как обучение с подкреплением, логико-комбинаторное распознавание и рекуррентные нейронные сети.

**Ключевые слова**: классификация, логико-комбинаторный подход, обучение с учителем с подкреплением.